\documentclass[aps,pra,superscriptaddress,amsmath,amssymb,preprintnumbers,floatfix,showpacs,11pt]{revtex4}
\usepackage{amssymb}
\usepackage{epsfig}

\begin{document}
\title{Relationship between the atomic inversion and  Wigner function for
multimode multiphoton Jaynes-Cummings model}
\author{Faisal A. A. El-Orany }
\email{el_orany@yahoo.com} \affiliation{ Department of Mathematics
and Computer Science, Faculty of Science, Suez Canal University
41522, Ismailia, Egypt }

\begin{abstract}
In this paper we consider multimode multiphoton Jaynes-Cummings
model, which consists of a two-level atom, initially prepared in
an excited atomic state, interacting with $N$ modes of
electromagnetic field prepared in general pure quantum  states.
For this system we show that under certain conditions the
evolution   of the Wigner function at the phase space origin
provides direct
 information on  the corresponding atomic inversion.
This relation is also valid even if the system includes Kerr-like
nonlinearity, Stark shift effect, different types of the initial
atomic state as well as moving atom. Furthermore, based on this
fact we discuss for the single-mode case the possibility of
detecting  the atomic inversion  by means of techniques similar to
those used for  Wigner function.

\end{abstract}
  \maketitle

\section{Introduction}

 Jaynes-Cummings model (JCM) \cite{jay1} has continued to be a subject
of not only theoretical studies but also experimental investigation
(, e.g. see \cite{remp}).   This model in the simplest form is described as
a two-level-atomic system
interacting with an electromagnetic field (for a review
see, e.g. \cite{stenh}). Many of quantum features of the JCM have been predicted
and observed: among the most well known is the revival-collapse phenomenon
(RCP) of the atomic inversion $\langle \hat{\sigma}_{z}(T)\rangle$
\cite{allen,{eber}}.
RCP is arised from the presence of multiple exchange of photons between
the radiating atom and the cavity mode.
Observation of RCP for $\langle \hat{\sigma}_{z}(T)\rangle$ has been
performed using the one-atom mazer \cite{remp}, which is more sophisticated than the
dynamics of the JCM.
In the same respect  it has been shown  that the measured probability of atomic
inversion for specific interaction times turns out to be the
symmetrically ordered characteristic function \cite{kim}. This
scheme is closely related to the nonlinear atomic homodyne detection
\cite{marin} in
which an atom is coupled to two modes of the field, one acting as the
signal mode and the other as the local oscillator mode.
Furthermore, in \cite{Ever} simple
scheme, which can slow down the usual exponential decay of upper state
population in an atomic two-level system considerably based on an additional
intense field with frequency lower than the total decay width of the
atom, is given.

Multiphoton single-mode JCM has taken a considerable
interest  in the literature, e.g. \cite{Suk, {Men},{Maq}, {El-O}}.
For instance, for this model the Heisenberg's equations of motion for the atomic energy
operator have been exactly sloved \cite{Suk}. Its  phase variance can
exhibit RCP about the long-time behavior \cite{Men}.  Moreover,
the investigation of this model against squeezed light has shown that
the atomic inversion can display RCP for general
squeezed input but not for squeezed vacuum \cite{Maq}. The analysis
of the the model against superposition of squeezed displaced number
states, i.e.  the most general case, is given in \cite{El-O}.
The multimode version of the JCM has been investigated, in particular, the
two-mode JCM, e.g. \cite{Gou,{card}}.
The most important result related to the two-mode JCM is that
the atomic inversion exhibits revival-collapse pattern as well as
secondary revivals, which are independent of the
intensities of the initial modes \cite{card}.
Moreover, we can mention \cite{Rosen}
  in which the hamiltonian
for the multiphoton multimode JCM has been derived from the
first-principle. Nevertheless, the generalizations of the JCM as a nonlinear
version in both bosonic and fermonic variables is given in \cite{Koch},
where the exact wavefunction and energy levels are calculated.

Quasiprobability distribution functions are very useful tools in quantum
mechanics since
 they can be used in the calculation of the correlation function of
 operators as  classical-like integrals and in the
 transition to the classical physics.
  There are three types of such functions, namely,
the Wigner $W$, the Husimi $Q$ and the Glauber $P$ functions \cite{wign}.
These functions are not real probability function owing to the
position-momentum  uncertainty principle.
Actually, the $W$ function plays an exceptional role among all quasiprobability
distributions for several reasons: It contains
complete information about the state of the system (, i.e. it carries the
same information as the density operator or as the corresponding wave
function). It  provides proper marginal distributions for
individual phase-space variables.
It can be used to evaluate the symmetrically-ordered
moments for the operators of the system.
It is  sensitive to the interference
in phase space and consequently it provides a clear prediction to  the possible
occurrence of the  nonclassical effects of the quantum mechanical system.
In this respect, the $W$ function can be used to analyse the decoherence of the quantum
 system, i.e.
the process that limits the appearance of quantum effects and turns them
into classical phenomena \cite{El-O1,{El-O2}}. It is worth mentioning
that the decoherence  is useful
for applications which require keeping coherence in mesoscopic or macroscopic
systems such as quantum computation \cite{EKer1}.
Finally, the $W$ function can be determined from the knowledge
of the complete set of moments of system operators \cite{knight1}.

 For the single-mode JCM with field prepared initially in coherent light
 it has been shown  that there is a relation between the behavior
of the $Q$  distribution function and the occurrence of the RCP in $\langle \hat{\sigma}_{z}(T)\rangle$
 \cite{ris1,{ris2},{ris3},{ris4}}.
 For instance, the collapse of the Rabi oscillations
in the evolution of $\langle \hat{\sigma}_{z}(T)\rangle$
is reflected in the behavior of the $Q$ function as the splitting of
the initial shifted Gaussian distribution into two
distributions, which counter-rotate on a circle in the complex plane of
the distribution.
However, the revivals in   $\langle \hat{\sigma}_{z}(T)\rangle$
correspond to
the collision of the two  peaks of the $Q$ function to produce
a single-peak distribution, which is similar to the initial one.
It is worthwhile mentioning that such relation between the $Q$ function and
$\langle \hat{\sigma}_{z}(T)\rangle$  for
JCM is remarkable only when the amplitude of the initial coherent light
is very large. Additionally, the comparison between
 $\langle \hat{\sigma}_{z}(T)\rangle$ and the $Q$ function has to be
performed at the same specific values of the interaction time.
In this paper we give a new relation, which shows
that the information stored in
$\langle \hat{\sigma}_{z}(T)\rangle$ can be obtained from the evolution
of the $W$ function at the phase space origin (WOP). This relation depends on
both the type of the initial state  of the optical cavity field, the
values of the transition parameters
and the number of modes interacting with the two-level atom.
The motivation of the work is two-fold:

\noindent (i) $\langle \hat{\sigma}_{z}(T)\rangle$ can be measured using techniques
similar to those used for the $W$ function.\newline
(ii) $\langle \hat{\sigma}_{z}(T)\rangle$ can be used to provide information on the nonclassicality
of the bosonic system.\newline
Actually, these are novel results.

In this paper we consider the
interaction of  multiphoton $N$ modes of the electromagnetic field
with a two-level atom in terms of
the multimode multiphoton Jaynes-Cummings model JCM. The hamiltonian
 controlling the system  is given in
 the framework of rotating wave approximation.
We also consider the optical cavity modes are
initially prepared in general pure quantum states.
For this system we seek the relation between the evolution
of $\langle \hat{\sigma}_{z}(T)\rangle$ and WOP.
This will be done in the following organization:
In section 2 we give the basic
relations and
equations used throughout the paper.
In section 3 we discuss the main results as well as we  shed the light
on how one can measure
$\langle \hat{\sigma}_{z}(T)\rangle$ using techniques similar to those used for
the $W$ function.
Conclusions and remarks are summarized in section 4.
\section{Basic relations and equations}
In this section we give the basic relations and equations, which enable
us to justify the relationship between
$\langle \hat{\sigma}_{z}(T)\rangle$ and the $W$ function for JCM.
Firstly, within the dipole and  rotating wave approximation (RWA) the
general form of an idealized hamiltonian, which describes
the interaction of  multiphoton $N$ modes cavity field
with a two-level atom (JCM) is \cite{Rosen,{mir}}
\begin{equation}
\frac{\hat{H}}{\hbar}=\sum\limits_{j=1}^{N}\omega_{j}\hat{a}_{j}^{\dagger}
\hat{a}_{j}+
\omega_{a}\hat{\sigma}_{z}+
\lambda (\hat{\sigma}_{+}\prod\limits_{j=1}^{N}\hat{a}^{k_{j}}_{j}+
\hat{\sigma}_{-}\prod\limits_{j=1}^{N}\hat{a}^{\dagger k_{j}}_{j}),
 \label{6}
\end{equation}
where the $j$th mode is designated by
$\hat{a}_{j}\quad (\hat{a}_{j}^{\dagger}$) the usual photon
annihilation (creation) operator,
the frequency $\omega_{j}$ and the transition parameter $k_{j}$.
$\hat{\sigma}_{\pm}$ and $\hat{\sigma}_{z}$ are the Pauli spin
operators describing the atomic system, $\omega_{a}$ is
 the atomic transition frequency and  $\lambda$ is the atom-field coupling
constant.
The hamiltonian can be written as the sum of the two operators:
\begin{eqnarray}
\begin{array}{rl}
\hat{C}_{1}=\epsilon_{1}\hat{\sigma}_{z}+\sum\limits_{j=1}^{N}\omega_{j}
\hat{a}^{\dagger}_{j}\hat{a}_{j}, \\
\hat{C}_{2}=
\triangle\hat{\sigma}_{z}+
\lambda (\hat{\sigma}_{+}\prod\limits_{j=1}^{N}\hat{a}^{k_{j}}_{j}+
\hat{\sigma}_{-}\prod\limits_{j=1}^{N}\hat{a}^{\dagger k_{j}}_{j}), \label{7}
\end{array}
\end{eqnarray}
where
\begin{equation}
\epsilon_{1}=\sum\limits_{j=1}^{N}k_{j}\omega_{j},\quad
\triangle=\omega_{a}-\epsilon_{1} \label{7as}
\end{equation}
and $\triangle$ is the detuning parameter.
Based on the standard commutation rules for the bosonic and Pauli
operators,
it is easy to prove that $\hat{C}_{1}$ and $\hat{C}_{2}$ are constants
of motion and also they  commute with each other.
In the interaction
picture  the unitary evolution operator takes the form
\begin{eqnarray}
\begin{array}{lr}
\hat{U}_{I}(T,0)=\exp(-i\frac{T}{\lambda}\hat{C}_{2})
\\
\\
=\sum\limits_{n=0}^{\infty}\frac{(-i\frac{T}{\lambda}\hat{C}_{2})^{n}}
{n!}\\
\\
=\sum\limits_{n=0}^{\infty}\frac{(-i\frac{T}{\lambda}\hat{C}_{2})^{2n}}
{(2n)!}+
\sum\limits_{n=0}^{\infty}\frac{(-i\frac{T}{\lambda}\hat{C}_{2})^{2n+1}}
{(2n+1)!}\\
\\
=\sum\limits_{n=0}^{\infty}\frac{(-1)^{n}(T\hat{\nu})^{2n}}
{(2n)!}-\frac{i}{\lambda\hat{\nu}}
\sum\limits_{n=0}^{\infty}\frac{(-1)^{n}(T\hat{\nu})^{2n+1}}
{(2n+1)!}\hat{C}_{2}\\
\\
=\cos (T\hat{\nu})-i\frac{\sin (T\hat{\nu})}{\lambda\hat{\nu}}\hat{C}_{2},\label{8}
\end{array}
\end{eqnarray}
where
\begin{equation}
T=\lambda t,\qquad \hat{\nu}^{2}=(\frac{\triangle}{\lambda})^{2}+
\hat{\sigma}_{-}\hat{\sigma}_{+}
\prod\limits_{j=1}^{N}\hat{a}^{\dagger k_{j}}_{j}
\hat{a}^{ k_{j}}_{j}
+ \hat{\sigma}_{+}\hat{\sigma}_{-}
\prod\limits_{j=1}^{N}\hat{a}^{ k_{j}}_{j}
\hat{a}^{\dagger k_{j}}_{j}. \label{9}
\end{equation}

For making the analysis quite general, we assume that
the $j$th mode is  initially prepared  in a general pure quantum state given
by
\begin{equation}
|\psi_{j}(0)\rangle=\sum\limits_{n_{j}=0}^{\infty}C^{(j)}_{n_{j}}|n_{j}\rangle, \label{1}
\end{equation}
where $C^{(j)}_{n_{j}}$ represents the probability amplitude for the
state under consideration such that
$\sum\limits_{n_{j}=0}^{\infty}|C^{(j)}_{n_{j}}|^{2}=1$.
We suppose that  the atom is initially prepared  in the
 excited  atomic state $|+\rangle$.
Therefore, the  initial state of the atom-field system  can be expressed as
\begin{eqnarray}
\begin{array}{lr}
|\Psi (0)\rangle=|\psi_{1} (0)\rangle\bigotimes
|\psi_{2} (0)\rangle\bigotimes\cdots \bigotimes
|\psi_{N} (0)\rangle\bigotimes
 |+\rangle\\
 \\
  =
  \sum\limits_{\underline{n}=\underline{0}}^{\infty}
  F(n_{1},n_{2},\cdots,n_{N})
  |+,n_{1},n_{2},\cdots,n_{N}\rangle,
  \label{3}
  \end{array}
\end{eqnarray}
where the vector notation in the index means that we have $N$ summations, i.e.
$\underline{n}\equiv (n_{1},n_{2},\cdots,n_{N})$, and the distribution
$F(n_{1},n_{2},\cdots,n_{N})$ reads
\begin{equation}
F(n_{1},n_{2},\cdots,n_{N})=\prod_{j=1}^{N}C^{(j)}_{n_{j}}.\label{3a}
\end{equation}

From (\ref{8}) and (\ref{3})  one can easily obtain
 the  dynamical wave function for the system in the interaction picture as
\begin{eqnarray}
\begin{array}{lr}
|\Psi(T)\rangle=\hat{U}_{I}(T,0)|\Psi(0)\rangle\\
\\
=  \sum\limits_{\underline{n}=\underline{0}}^{\infty}
  F(n_{1},n_{2},\cdots,n_{N})
  \Bigl[G_{1}(n_{1},n_{2},\cdots,n_{N},T)
  |+,n_{1},n_{2},\cdots,n_{N}\rangle \\
  \\
-i  G_{2}(n_{1},n_{2},\cdots,n_{N},T)|-,n_{1}+k_{1},\cdots,n_{N}+k_{N}\rangle
\Bigr], \label{10}
\end{array}
\end{eqnarray}
where
\begin{eqnarray}
\begin{array}{lr}
h(n_{1}, n_{2},\cdots ,n_{N};k_{1}, k_{2},\cdots ,k_{N})
=\prod\limits_{j=1}^{N}\frac{(n_{j}+k_{j})!}{n_{j}!}
+(\frac{\triangle}{\lambda})^{2},\\
\\
G_{1}(n_{1},n_{2},\cdots,n_{N},T)=
\cos\left(T\sqrt{h(n_{1}, n_{2},\cdots ,n_{N};k_{1}, k_{2},\cdots ,k_{N})}\right)
\\
\\
  -i\frac{\triangle}{\lambda}\frac{\sin\left(T\sqrt{
  h(n_{1}, n_{2},\cdots ,n_{N};k_{1}, k_{2},\cdots ,k_{N})}\right)}
{\sqrt{h(n_{1}, n_{2},\cdots ,n_{N};k_{1}, k_{2},\cdots ,k_{N})}},\\
\\
G_{2}(n_{1},n_{2},\cdots,n_{N},T)=
-\frac{\sin\left(T\sqrt{h(n_{1}, n_{2},\cdots ,n_{N};k_{1}, k_{2},\cdots ,k_{N})}\right)}
{\sqrt{h(n_{1}, n_{2},\cdots ,n_{N};k_{1}, k_{2},\cdots ,k_{N})}}
\sqrt{\prod\limits_{j=1}^{N}\frac{(n_{j}+k_{j})!}{n_{j}!}}.
\label{11}
\end{array}
\end{eqnarray}
The atomic inversion associated with (\ref{10}) is
\begin{equation}
 \langle \sigma_{z}(T)\rangle=
  \sum\limits_{\underline{n}=\underline{0}}^{\infty}
 |F(n_{1},n_{2},\cdots,n_{N})|^{2}[
 |G_{1}(n_{1},n_{2},\cdots,n_{N},T)|^{2}
 -|G_{2}(n_{1},n_{2},\cdots,n_{N},T)|^{2}].
\label{12}
\end{equation}

For reasons that will be clear shortly we write down the different forms for
the $W$ function.
The basis of  the $W$ function for any quantum mechanical system
 is the $W$ function of the number state $|n\rangle$ having the form
\begin{equation}
W_{n}(q,p)=\frac{(-1)^{n}}{\pi}\exp
(-q^{2}-p^{2})
{\rm L}_{n}(2q^{2}+2p^{2}), \label{1ac}
\end{equation}
where ${\rm L}_{n}(.)$ is the Laguerre polynomial of order $n$.
Also the marginal position probability distribution
for the number state $|n\rangle$ can be obtained from
 (\ref{1ac})  as
\begin{eqnarray}
\begin{array}{lr}
P(q)=\int\limits_{-\infty}^{\infty}W_{n}(q,p)dp\\
\\
=\frac{1}{\sqrt{\pi}}\frac{{\rm
H}^{2}_{n}(q)}{2^{n}n!}\exp(-\frac{q^{2}}{2}),\label{1ad}
\end{array}
\end{eqnarray}
where ${\rm H}_{n}(.)$ is the Hermite polynomial of order $n$. The
corresponding form of the marginal momentum probability distribution is
the same as (\ref{1ad}) but $q$ should be replaced by $p$.
The $N$-mode dynamical $W$ function can be defined
up to a constant prefactor as
\cite{wign}
\begin{equation}
W(\underline{\beta},T)=
{\rm Tr}\left[\hat{\rho}(T)
\hat{D}(\underline{\beta})\exp\left(i\pi\sum\limits_{j=1}^{N}
\hat{a}^{\dagger}_{j}\hat{a}_{j}\right)\hat{D}^{-1}(\underline{\beta})\right],
\label{16}
\end{equation}
where $\underline{\beta}=(\beta_{1}, \beta_{2},\cdots,\beta_{N})=(q_{1},q_2,
\cdots,q_N;p_1,p_2,\cdots,p_N)$ since
$\beta_{j}=q_{j}+ip_{j}$.
 $\hat{\rho}(T)$ is the density matrix for the system under consideration
 and  $\hat{D}(\underline{\beta})$ is the multimode
displacement operator having the form
\begin{equation}
\hat{D}(\underline{\beta})
=\exp\left[\sum\limits_{j=1}^{N}(\hat{a}^{\dagger}_{j}\beta_{j}-\hat{a}_{j}\beta^{*}_{j})\right].
\label{15}
\end{equation}
At the phase space origin (, i.e. $\underline{\beta}=\underline{0}$)
the formula (\ref{16}) reduces to
\begin{equation}
W(\underline{0},T)=
{\rm Tr}\left[\hat{\rho}(T)\exp\left(i\pi\sum\limits_{j=1}^{N}
\hat{a}^{\dagger}_{j}\hat{a}_{j}\right)
\right].
\label{17}
\end{equation}
Formula (\ref{17}) indicates that the main contribution for the $W$ function
at the phase space origin is resulted from the diagonal part of the
density matrix of the quantum mechanical system.  Comparing this situation with
that of the atomic inversion one can conclude that there is a clear
relationship between the evolution of WOP and the corresponding atomic
inversion.

Now on substituting (\ref{10})
into (\ref{17}) and carrying out the expectation value we arrive at

\begin{eqnarray}
\begin{array}{lr}
W(\underline{0},T)=
  \sum\limits_{\underline{n}=\underline{0}}^{\infty}
  |F(n_{1},n_{2},\cdots,n_{N})|^{2} (-1)^{n_{1}+n_{2}+\cdots+n_{N}}
    \Bigl\{
|G_{1}(n_{1},n_{2},\cdots,n_{N},T)|^{2}\\
\\
  +(-1)^{k_{1}+\cdots+k_{N}}|G_{2}(n_{1},n_{2},\cdots,n_{N},T)|^{2}
\Bigr\}. \label{18}
\end{array}
\end{eqnarray}
Expression (\ref{18}) and its consequences are the main results of the paper.
Specifically, we show that for particular types of the initial states  and
particular values of the transition parameters $k_{j}$, expression
(\ref{18}) coincides with that of the corresponding atomic inversion.
This  will be discussed  in the following section.

We proceed by  connecting the present results with those of the homodyne
tomography technique. Therefore, we give the mathematical relation  between the $N$-mode $W$ function and the
corresponding distribution function
$pr(q_{1},\cdots,q_{N},\theta_{1},\cdots,\theta_{N},T)$ (, i.e. Radon transformation).
Such relation is just the generalization of  the single-mode case and can be
expressed as
\begin{eqnarray}
\begin{array}{lr}
pr(q_{1},\cdots,q_{N},\theta_{1},\cdots,\theta_{N},T)
=\int\limits_{-\infty}^{\infty}dp_{1}\cdots\int\limits_{-\infty}^{\infty}dp_{N}\\
\\
\times W(
q_{1}\cos\theta_{1}-p_{1}\sin\theta_{1},q_{1}\sin\theta_{1}+p_{1}\cos\theta_{1},
\cdots ,q_{N}\cos\theta_{N}-p_{N}\sin\theta_{N},q_{N}\sin\theta_{N}+p_{N}\cos\theta_{N},T). \label{1aa}
\end{array}
\end{eqnarray}
In (\ref{1aa}) we have assumed  that $N$ modes can be delivered to $N$ separate
ideal balance homodyne detectors.
At the phase space origin, i.e. $q_j=0, \theta_j=0, j=1,\cdots N$, the formula (\ref{1aa}) becomes phase independent
and reduces to
\begin{equation}
pr(\underline{0},T)
=\int\limits_{-\infty}^{\infty}dp_{1}\cdots
\int\limits_{-\infty}^{\infty}dp_{N}
W(0,p_{1},0,p_2,\cdots ,0,p_{N},T), \label{1ab}
\end{equation}
where $W(0,p_{1},0,p_2,\cdots ,0,p_{N},T)$ is the diagonal part of the $W$ function, which
is phase independent. On using (\ref{1ac}) and
 (\ref{1ad}) one can easily deduce
 $pr(\underline{0},T)(=P(\underline{0},T))$ given by (\ref{1ab})
 for the state vector (\ref{10})  as

\begin{eqnarray}
\begin{array}{lr}
P(\underline{0},T)=
  \sum\limits_{\underline{n}=\underline{0}}^{\infty}
  |F(n_{1},n_{2},\cdots,n_{N})|^{2} \Bigl\{
|G_{1}(n_{1},n_{2},\cdots,n_{N},T)|^{2}
\prod\limits_{j=1}^{N} \frac{{\rm H}^{2}_{n_{j}}(0)}{2^{n_{j}}n_{j}!}
 \\
 \\
  +
|G_{2}(n_{1},n_{2},\cdots,n_{N},T)|^{2} \prod\limits_{j=1}^{N}
\frac{{\rm H}^{2}_{n_{j}+k_{j}}(0)}{2^{n_{j}+k_{j}}(n_{j}+k_{j})!}
\Bigr\}. \label{1af}
\end{array}
\end{eqnarray}

\section{Main results}
In this section we discuss two issues:
(i) We investigate the results by
making a comparative study among the behavior of
$\langle \hat{\sigma}_{z}(T)\rangle, W(\underline{0},T)$ and $P(\underline{0},T)$,
for the system under consideration.
(ii) We argue how one can  measure the
atomic inversion via techniques similar to those used for the $W$ function.
\subsection{Investigation of the  results}
As we mentioned above we investigate the behavior of the quantities
$W(\underline{0},T)$ and   $P(\underline{0},T)$, and then compare such behavior
with that of the corresponding $\langle \hat{\sigma}_{z}(T)\rangle$.

We start the discussion with $W(\underline{0},T)$, which is given by (\ref{18}).
It is obvious that
  when $n_{1}+n_{2}+\cdots+n_{N}$ is even and
$k_{1}+k_{2}+\cdots+k_{N}$ is odd
$W(\underline{0},T)$ is identical with the atomic
inversion of the system (c.f. (\ref{12})). In other words,
the atomic inversion can be used to provide information
on the nonclassicality of the dynamical bosonic system.
For instance, when the evolution of
$\langle\hat{\sigma}_{z}(T)\rangle$ displays negative values,
the JCM can exhibit  nonclassical effects.
 Nevertheless, when
$k_{1}+k_{2}+\cdots+k_{N}$ is even number
regardless of the value of $n_{1}+n_{2}+\cdots+n_{N}$, the
 expression (\ref{18})
becomes time independent (, i.e. $W(\underline{0},T)$ is localized)
 and can be factorized in the following sense
\begin{equation}
W(\underline{0},T)=\prod_{j=1}^{N}
W_{j}(0,0)
, \label{19}
\end{equation}
 where $W_{j}(0,0)$ is the initial value of the $W$ function of the
 $j$th mode at the phase space origin having the form
\begin{equation}
W_{j}(0,0)=\sum\limits_{n_j=0}^{\infty}(-1)^{n_j}|C^{(j)}_{n_j}|^{2}.
 \label{191}
\end{equation}
The expression (\ref{19}) can be obtained, e.g., when
the number of  modes $N$ is even and
the transition parameters
are symmetric, i.e. $k_{1}=k_{2}=\cdots =k_{N}$.
Moreover, expression (\ref{19}) indicates that if the initial $W$ function
 of only one of the modes has a negative value at the phase space origin
 whereas those of the others are positive,
 the system can provide nonclassical effects.
This is a sufficient but not necessary condition.

\begin{figure}
  \vspace{0cm}
\centerline{\epsfxsize=16cm \epsfbox{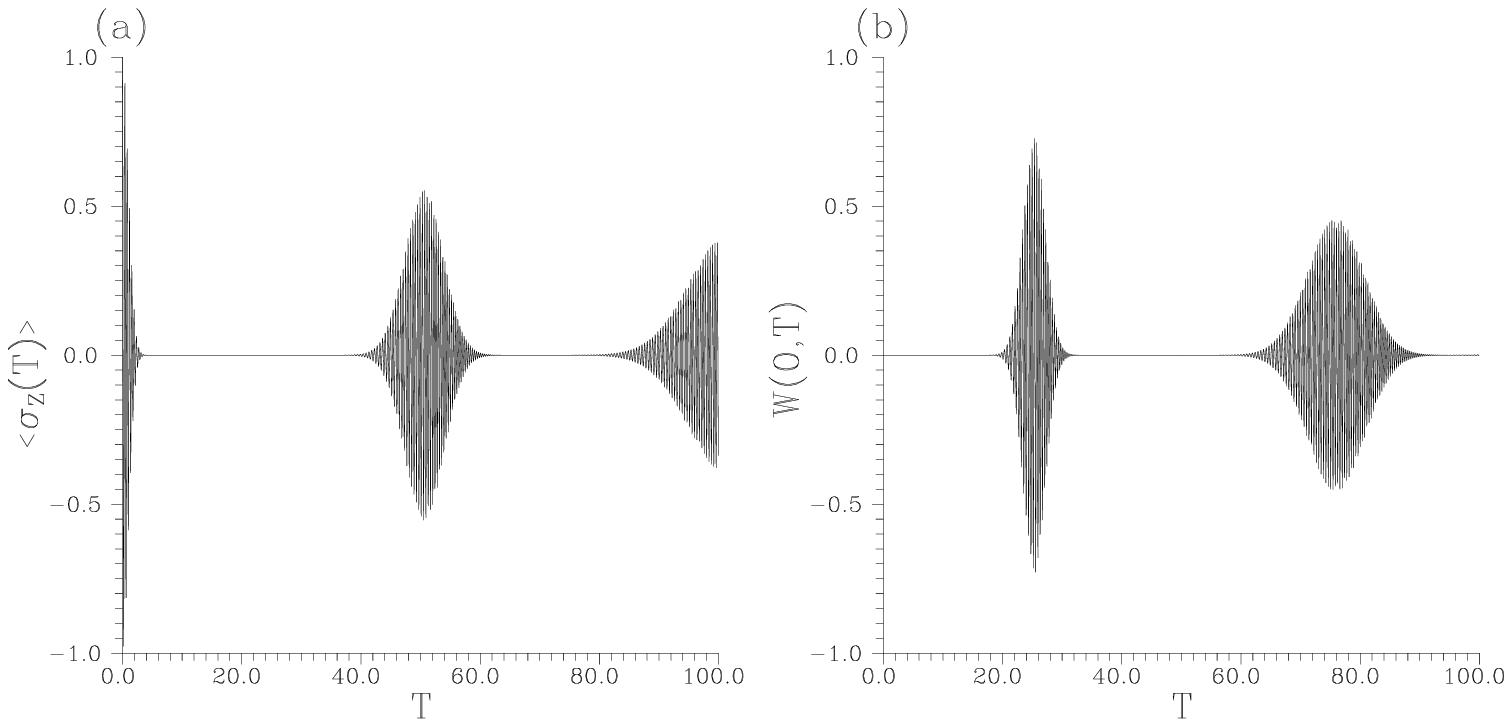} } \vspace{.1cm}
\caption{ The evolution of the atomic inversion
$\langle\hat{\sigma}_{z}(T)\rangle$ (a) and the $W$ function
$W(0,T)$  (b) against the scaled time $T$ for the single-mode case
with $k_{1}=1$ and when the field and atom are initially prepared
in the coherent state (with $|\alpha|=8$) and atomic excited
state, respectively.}
\end{figure}

Now we give a closer look at the behavior of the $W(0,T)$ for the
single-mode case, i.e. $N=1$. In this case expression (\ref{18})
reduces to
\begin{equation}
 W(0,T)=
  \sum\limits_{n_{1}=0}^{\infty}  |C^{(1)}_{n_{1}}|^{2} (-1)^{n_{1}}
  \Bigl\{
  |G_{1}(n_{1},T)|^{2}+
(-1)^{k_{1}}|G_{2}(n_{1},T)|^{2}
\Bigr\}. \label{20}
\end{equation}
For odd transition parameter and initial even (odd)  parity states,
 e.g. even (odd) coherent states,
  (\ref{20}) gives
\begin{equation}
W_{\pm}(0,T)=
\pm\langle\sigma_{z}(T)\rangle, \label{21}
\end{equation}
where "+"  and "-" signs denote even and odd parity states, respectively.
For $\triangle=0$ expression (\ref{21}) indicates that when
the initial intensity of the radiation field is  weak
 the system can exhibit nonclassical effects periodically.
Nevertheless,  in the strong-intensity regime one has
$W(0,T)\simeq 0 \quad (\neq 0)$, which is associated with the occurrence of the
collapse (revival) in $\langle \hat{\sigma}_{z}(T)\rangle$. Consequently
the $W$ function exhibits nonclassical interference at the phase space origin
only  in the course of the revival times, i.e.
 the nonclassical effects most probable occur in the course of
the revival time. However for the non-parity states the
locations (in the interaction time domain) of collapses and revivals occurring
in  $W(0,T)$ are interchanged compared to those in $\langle\hat{\sigma}_{z}(T)\rangle$.
This agrees with the fact that the JCM generates Schr\"{o}dinger-cat states
in the course of the collapse time \cite{El-O}.
We proceed by investigating the behavior of the $W(0,T)$
 for the standard JCM, i.e. $k_{1}=1, \triangle=0$ and the field
is initially  prepared in coherent light with amplitude $|\alpha|$.
In this case (\ref{20}) reduces to

\begin{eqnarray}
\begin{array}{lr}
W(0,T)=
\exp(-|\alpha|^{2})
  \sum\limits_{n_{1}=0}^{\infty}  \frac{|\alpha|^{2n_{1}}}{n_{1}!}
  (-1)^{n_{1} }
\cos (2T\sqrt{n_{1}+1})\\
\\
=
\exp(-|\alpha|^{2})
  \sum\limits_{n_{1}=0}^{\infty}  \frac{|\alpha|^{2n_{1}}}{n_{1}!}
\cos(2T\sqrt{n_{1}+1}+n_{1}\pi). \label{22}
\end{array}
\end{eqnarray}
Expression (\ref{22}) is identical with that of the corresponding atomic
inversion but with additional factor, which is $(-1)^{n_{1}}$.
This factor is responsible for the interchange of
the "locations" of collapses and
revivals occurring in $W(0,T)$ compared to those exhibited in
$\langle \hat{\sigma}_{z}(T)\rangle$, as we mentioned above.
This is  remarkable in Figs. 1(a) and (b) where we have plotted
$\langle \hat{\sigma}_{z}(T)\rangle$ and $W(0,T)$, respectively,
for given values of the interaction parameters.

The evolution of the atomic inversion
$\langle\hat{\sigma}_{z}(T)\rangle$ (a) and thethe $W$ function
$W(0,T)$  (b) against the scaled time $T$ for the single-mode case
with $k_{1}=1$ and when the field and atom are initially prepared
in the coherent state (with $|\alpha|=8$) and atomic excited
state, respectively.
\begin{figure}
  \vspace{0cm}
\centerline{\epsfxsize=12cm \epsfbox{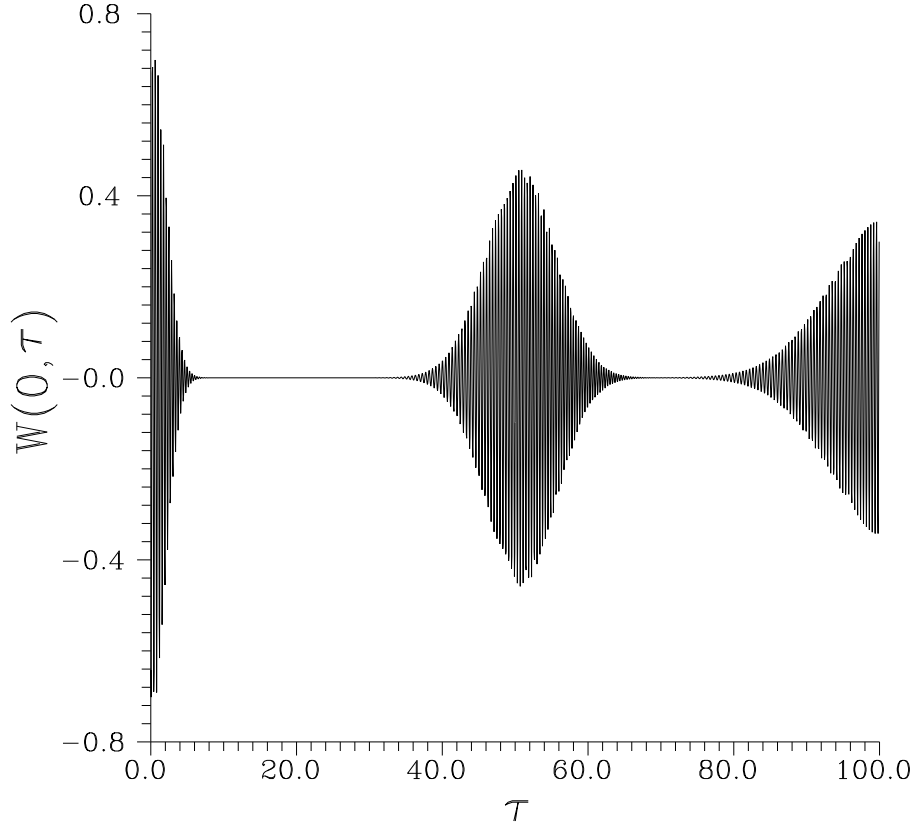} } \vspace{.1cm}
\caption{ The evolution of the $W(0,\tau)$ against the
shifted-scaled time $\tau$ for the same situation as in Fig.
1(b).}
\end{figure}

Also this can be emphasized by deducing the asymptotic form for
(\ref{22}) in the strong-intensity regime (, i.e. $|\alpha|$ is
large). By means of  the harmonic approximation technique
\cite{rice} (see equation (1a) in the appendix) and after
straightforward calculations (\ref{22}) can be expressed as
\begin{equation}
W(0,T)=
\exp\left[-2\bar{n}\cos^{2}(\frac{T}{2\bar{n}})\right]
\cos\left[T(\bar{n}+\frac{1}{\bar{n}})-\bar{n}\sin(\frac{T}{\bar{n}})\right]
, \label{24}
\end{equation}
where $\bar{n}=|\alpha|$. Expression (\ref{24}) is similar to that of
$\langle\hat{\sigma}_{z}(T)\rangle$  except
$\cos(.)$ in the exponent should be  replaced by $\sin(.)$.
As a result of this fact the envelope function in (\ref{24})
gives its maximum value at
 $T=\pi\bar{n}$, whereas, that of
$\langle\hat{\sigma}_{z}(T)\rangle$ is maximum at $T=2\pi\bar{n}$.
From these arguments and  information displayed in Figs. 1
one can conclude that  $W(0,T)$ can  give similar
 information on the corresponding atomic inversion provided that the interaction
 time $T$ is replaced by
  $\tau\equiv T+\pi \bar{n}$. The behavior associated with this situation is
 given in Fig. 2. Comparison between Fig. 1(a) and  Fig. 2 is instructive.

\begin{figure}
  \vspace{0cm}
\centerline{\epsfxsize=16cm \epsfbox{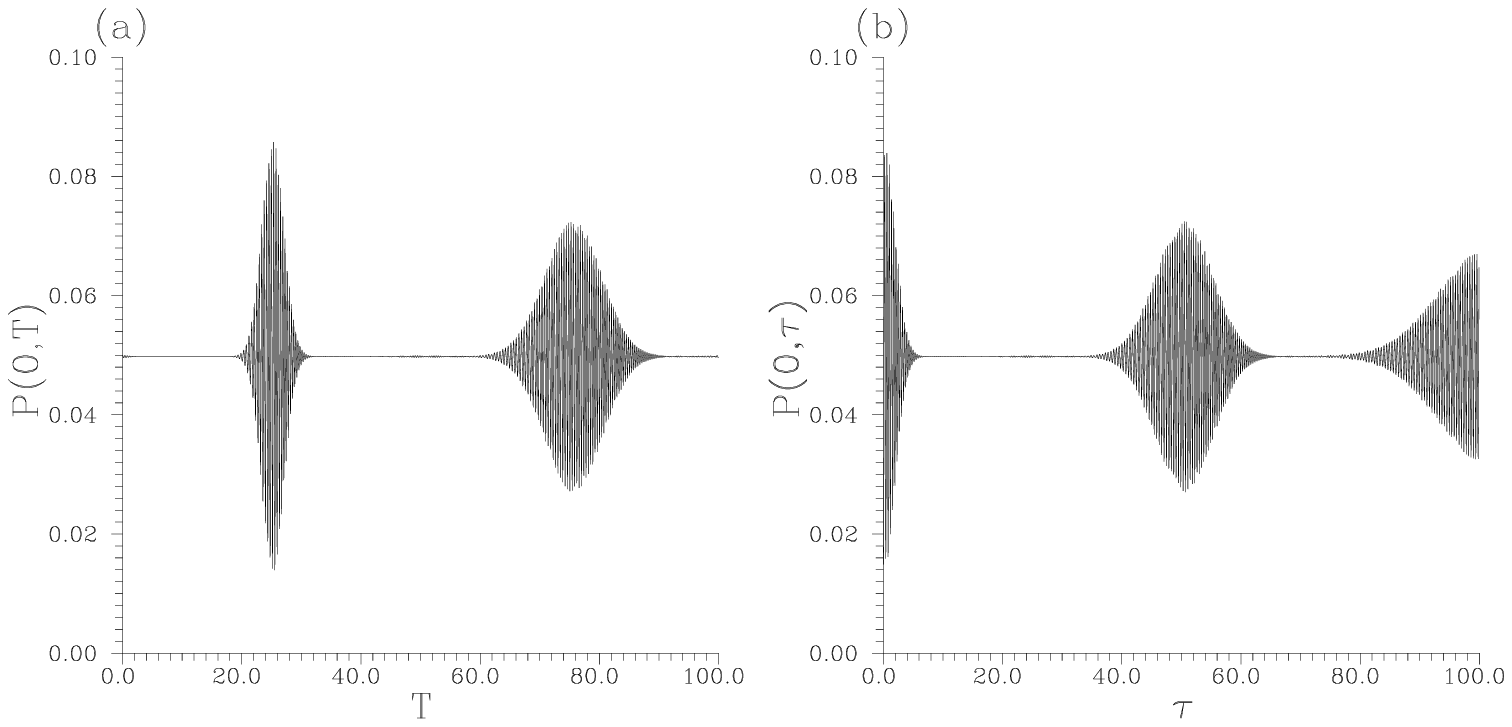} } \vspace{.1cm}
\caption{ The probability distribution measured using homodyne
tomography corresponding to the evolution of $W(0,T)$
 against both the scaled time $T$ (a) and shifted-scaled time
 $\tau$ (b) for the same situation as figure 1(b) and figure 2, respectively.}
\end{figure}

Now we turn the attention to the formula (\ref{1af}), which is related to
the homodyne tomography.
Firstly, it is worth reminding that the properties of
the Hermite polynomial provide ${\rm H}_{2n+1}(0)=0$ and
${\rm H}_{2n}(0)\neq 0$. These facts make the argument related to
$P(\underline{0},T)$   different from that given
for $W(\underline{0},T)$.
For instance, when one of the modes is initially prepared
in odd parity states
(, e.g. odd coherent states) and the associated transition parameter with this
mode is an even
number  then $P(\underline{0},T)=0$, however, this is not the case of
the corresponding $W$ function,
where $W(\underline{0},T)\neq 0$. Also one can easily recognized that
$P(\underline{0},T)\neq 0$ for different cases, e.g.,
when  all  modes are initially in even parity (non-parity)
states regardless of the values of $k_{j}$ or when all modes are initially
in odd parity states provided that the transition parameters are odd numbers.

Similar to the treatment given for the $W$ function
 we investigate $P(0,T)$ of the single-mode case when the field is initially
in coherent state, $k_{1}=1$ and $\triangle=0$. Therefore,
relation (\ref{1af}) can be expressed as

\begin{eqnarray}
\begin{array}{lr}
P(0,T)=\frac{1}{2}
  \sum\limits_{n_{1}=0}^{\infty}
  P(n_{1}) \Bigl\{
  \left[ \frac{{\rm H}^{2}_{n_{1}}(0)}{2^{n_{1}}n_{1}!}
 +
  \frac{{\rm H}^{2}_{n_{1}+1}(0)}{2^{n_{1}+1}(n_{1}+1)!}\right]\\
  \\
 +
  \left[ \frac{{\rm H}^{2}_{n_{1}}(0)}{2^{n_{1}}n_{1}!}-
  \frac{{\rm H}^{2}_{n_{1}+1}(0)}{2^{n_{1}+1}(n_{1}+1)!}\right]
 \cos(2T\sqrt{n_{1}+1})
\Bigr\}, \label{24a}
\end{array}
\end{eqnarray}
where $P(n_{1})$ is the photon number distribution for coherent states.
In the strong-intensity regime the asymptotic form for (\ref{24a}), which is
corresponding to (\ref{24}), is
\begin{eqnarray}
\begin{array}{lr}
P(0,T)=\frac{1}{2}\exp(-|\alpha|^{2})
 \Bigl\{
   {\rm I}_{0}(|\alpha|^{2})+
   {\rm I}_{1}(|\alpha|^{2})
    +
    \\
    \\
 {\rm Re}\Bigl\{\exp[iT(\bar{n}+\frac{1}{\bar{n}})]\left[
   {\rm I}_{0}(|\alpha|^{2}\exp(i\frac{T}{\bar{n}}))-
{\rm I}_{1}(|\alpha|^{2}\exp(i\frac{T}{\bar{n}})) \right]\Bigr\}
\Bigr\}, \label{24b}
\end{array}
\end{eqnarray}
where ${\rm I}_{0}(.)$
and ${\rm I}_{1}(.)$ are the modified Bessel functions of the first kind
of order zero and one, respectively. The derivation
for (\ref{24b}) is given in the appendix.
Expression (\ref{24b}) is periodic with period $2\pi \bar{n}$
where $\bar{n}$ is an integer.
Additionally, $P(0,T)$ gives its maximum values around $T=m\bar{n}\pi$ whenever
$m$ is an odd integer. We have plotted (\ref{24a}) in figures (3) against
 both the scaled time $T$ (a) and the shifted-scaled time
 $\tau$ (b) for the same situations as those for figure 1(b) and figure 2, respectively.
 Comparison between Fig. 1(b) and Fig. 3(a) as well as
 Fig. 2 and Fig. 3(b) shows that RCP in
$\langle\hat{\sigma}_{z}(T)\rangle$ can be observed via homodyne tomography.
Of course, the evolution of $\langle\hat{\sigma}_{z}(T)\rangle$
and $P(0,T)$ possess different scales.

\begin{figure}
  \vspace{0cm}
\centerline{\epsfxsize=12cm \epsfbox{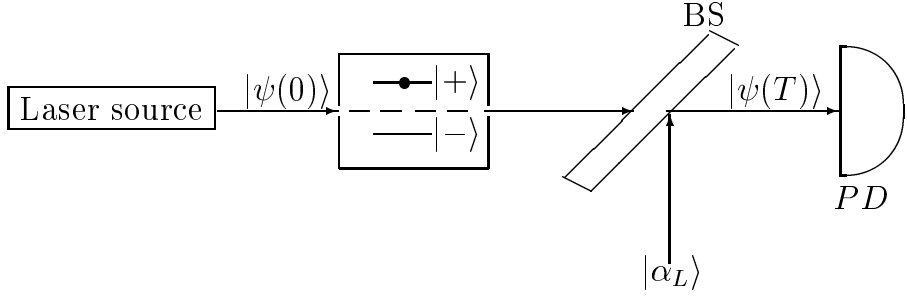} } \vspace{.1cm}
\caption{ The proposed setup for detecting
$\langle\hat{\sigma}_{z}(T)\rangle$ of the single-mode JCM when
the field is initially prepared in the coherent light pumped by
the laser source.  BS and PD denote  beam splitter and
photodetector, respectively. The two-level excited atom is
localized in the microcavity as indicated. The initial state,
detected state and local-oscillator state are denoted by
$|\psi(0)\rangle, |\psi(T)\rangle$ and $|\alpha_{L}\rangle$,
respectively.}
\end{figure}

This problem can be solved easily by using (\ref{24b}) and the
information displayed in Fig. 3(b). For instance, in the
strong-intensity regime we can adopt the following relation
\begin{equation}
\langle\hat{\sigma}_{z}(T)\rangle\equiv\frac{1}{P_{max}}[P(0,\tau)-P(0,0)],
\label{24c}
\end{equation}
where
\begin{eqnarray}
\begin{array}{lr}
P(0,0)=\exp(-|\alpha|^{2})  {\rm I}_{0}(|\alpha|^{2}),\\
\\
P_{max}=[{\rm I}_{0}(|\alpha|^{2})+
   {\rm I}_{1}(|\alpha|^{2})]\exp(-|\alpha|^{2}), \label{24d}
\end{array}
\end{eqnarray}
where the subscript $max$ stands for the maximum value of $P(0,T)$.
The explicit form for $P_{max}$ can be obtained by analysing
the behavior of $P(0,T)$ around $T=\pi\bar{n}$.
 Finally, the origin of the prefactor (, i.e. $1/P_{max}$)
 in (\ref{24c}) can be understood as follows.
$P(0,T)$ carries  complete information on
$\langle\hat{\sigma}_{z}(T)\rangle$ when it has not only similar behavior
but also  the same amplitude as $\langle\hat{\sigma}_{z}(T)\rangle$.
Thus we seek an amplification factor, say, $\mu$
(, i.e. $\mu P(0,T)\equiv\langle\hat{\sigma}_{z}(T)\rangle$)
such that
\begin{equation}
\mu
P_{max}\equiv |\langle\hat{\sigma}_{z}(T)\rangle_{max}|=1.
\label{24f}
\end{equation}
\subsection{Observation and measurement}
In  the first part of this section   we have shown that under certain conditions
 there is a direct relation between
$\langle\hat{\sigma}_{z}(T)\rangle$ and $W(0,T)$ for JCM. Such relation indicates
 that the atomic inversion can be detected via techniques similar to those
used for the $W$ function. This will be discussed in the
following, in particular, for the single-mode JCM.

As it is well known there are different schemes proposed for measuring
 the $W$ function, which are:
Photon counting experiment \cite{con}, using simple experiment
similar to that used  in cavity QED and
ion traps  \cite{ion1,{ion2}},
and tomographic reconstruction from data obtained
in homodyne measurements \cite{vog,{tom}}.
Here we argue how these techniques can be used for measuring
$\langle \hat{\sigma}_{z}(T)\rangle$.

 (i) Measurement of the $\langle \hat{\sigma}_{z}(T)\rangle$
using photon counting experiment:
Photon counting  method is based on the fact that the single-mode
$W$ function at the origin of
the  phase space  can  directly be measured by a
photodetector facing this mode \cite{con}. Hence
 $\langle\hat{\sigma}_{z}(T)\rangle$ can be detected via this technique
  in the following sense (see Fig. 4).
Generally, a mode prepared in a coherent state, which is pumped
by a laser source,  interacts firstly  with the two-level excited localized
atom---localized and/or very slow atom can be prepared by means of, e.g.,
laser-cooling technique \cite{laser}---hence the outgoing field is
superimposed by a strong
local-oscillator mode ($|\alpha_{L}\rangle$) via  beam splitter whose
 transmissivity  (reflectivity) is high (low). The measurement has to be
performed only
on one of the output ports of the beam splitter via the photodetector.
It is worthwhile mentioning that the one-port measurement for
the beam splitter can be performed using a conditional
measurement technique in which no photons are measured in the free
port, e.g., \cite{meas}.  We proceed by considering a perfect photodetector, i.e. its
efficiency is unity.  The probability $P(n_{1},T)$ of the registration
of $n_{1}$ photons at time $T$ in the detector is given by
\begin{equation}
P(n_{1},T)=\langle :\frac{\hat{\Lambda}^{n_{1}}}{n_{1}!}\exp (- \hat{\Lambda}):
\rangle,\label{27}
\end{equation}
where
$: :$ stands for normally-ordered operator, angle brackets mean
 expectation value (which is calculated in the framework of  Schr\"{o}dinger picture) and
$\hat{\Lambda}$ is the operator of the integrated flux of
light onto the surface  of the detector having the form
\begin{equation}
\hat{\Lambda}=
(\sqrt{\bar{\tau}}\hat{a}^{\dagger}-\sqrt{1-\bar{\tau}}
\alpha^{*}_{L})
(\sqrt{\bar{\tau}}\hat{a}-\sqrt{1-\bar{\tau}}
\alpha_{L}), \label{28}
\end{equation}
where $\bar{\tau}$ is  the transmissivity power of the beam splitter.
The count statistics determined in the experiment is used to compute the
following count generating function
\begin{eqnarray}
\begin{array}{lr}
G(\alpha,T)=\sum\limits_{n_{1}=0}^{\infty} (-1)^{n_{1}}P(n_{1},T)\\
\\
  =\langle:\exp(-2\hat{\Lambda}):\rangle. \label{29}
  \end{array}
  \end{eqnarray}
The second line in (\ref{29}) is obtained by substituting
 (\ref{27}) in the first line of  this equation.
 It is evident that for $\triangle=0$,
 $W(0,T)$ given by (\ref{20}) can be expressed in the form (\ref{29}),
 where, in this case,  we have
 \begin{equation}
 P(n_{1},T)=
  |C^{(1)}_{n_{1}}|^{2} \left[
  \cos^{2}(T\sqrt{h(n_{1},k_{1})})
  +\sin^{2}(T\sqrt{h(n_{1},-k_{1})})\right]. \label{31}
\end{equation}
We proceed from  the definition of the $W$ function
(\ref{17})
 and (\ref{29}) the
$W$ function for the detected mode is proportional to
$s=1-1/\bar{\tau}$ ordered quasidistribution function of the mode entering
the beam splitter (which is not included in Fig. 4) as
\begin{equation}
G(\alpha,T)=\frac{1}{\bar{\tau}}W\left(
\sqrt{\frac{1-\bar{\tau}}{\bar{\tau}}}\alpha_{L},
\frac{\bar{\tau}-1}{\bar{\tau}},T\right). \label{30}
\end{equation}
When the transmissivity of the beam splitter is near one the
 scanned quasidistribution is  $W(0,T)$, i.e.
 the atomic inversion is detected. Here we assume that the
 interaction time in the microcavity and the detection time in
 the photodetector are equal since  we are interested only in showing that
$\langle \hat{\sigma}_{z}(T)\rangle$ can be measured via photon counting
technique.  Nevertheless,
in the realistic situation there is a time delay between the interaction and
the detection processes. This problem can be solved by obtaining
information on both the detector-microcavity distance
and the velocity of the radiation field.

 (ii) Measurement of the $\langle \hat{\sigma}_{z}(T)\rangle$
using cavity QED:
In this technique, probes of the two-level atoms
interact dispersively with the field under consideration--the source of
 the field in this case is the JCM to
which we would like to measure $\langle \hat{\sigma}_{z}(T)\rangle$--causing
 a phase shift
to the atomic wave function, which is proportional to the photon number.
This phase shift can be revealed by  Ramsey atomic interferometer \cite{rams}.
Under certain conditions the final state of the atom measures the field
parity at the phase space origin, i.e. $W(0,T)$.
The experimental setup related to this proposal  is given in \cite{ion1} (see Fig. 1 in
\cite{ion1}) but with additional arrangements.
For instance, in this setup we have
 to block the microwave generator, which makes
  displacement for the
 field under consideration since the interest is focused only on
the behavior of the $W$ function at the phase space origin.

 (iii) Measurement of the $\langle \hat{\sigma}_{z}(T)\rangle$
using homodyne tomography arrangement:
Generally  single-mode homodyne tomography is based on
the set of distributions $pr(q_{1},\theta_{1})$ measured by homodyne
detection, i.e.
the field to be measured beats with the local oscillator in a homodyne
arrangement \cite{yuen,{leon}}. Once obtained $pr(q_{1},\theta_{1})$
the $W$ function can be reconstructed via inverse Radon transformation.
At the phase space origin this relation reads
\begin{equation}
W(0,T)=\frac{1}{4\pi}\int\limits_{-\infty}^{\infty}d\zeta
\int\limits_{-\infty}^{\infty}d\eta |\eta|
pr(\zeta,T) \exp(i\eta\zeta), \label{33}
\end{equation}
where  $pr(\zeta,T)=\langle \zeta|\hat{\rho}(T)| \zeta\rangle$.
The value of the relative phase
between the local oscillator and the signal field, which is assumed to be
the field outgoing from the JCM microcavity, is zero.
This can be arranged by moving a mirror on a piezoelectric translator
 \cite{smithey}.
It is worthwhile mentioning that
 $pr(\zeta,T)$  has been measured via this technique, e.g., in \cite{smithey}.
Also  investigation for the random-phase states  using homodyne tomography
technique is given in  \cite{leon1}.

Finally we conclude that the techniques (i) and (ii)
can lead to a direct  measurement of
$\langle \hat{\sigma}_{z}(T)\rangle$. Also, they do not involve
inversion algorithm and hence they should
be much less sensitive to experimental errors than the tomographical technique.
Furthermore, they
can be, in principle, applied to the atomic inversion of the
 multimode JCM where the attention
has to be focused on the
measurement of the $\langle \hat{\sigma}_{z}(T)\rangle$ in its entangled form.

\section{Conclusions and remarks}
In this paper we have discussed  the relation between both the
evolution of the atomic inversion and the
corresponding $W$ function for multimode multiphoton JCM. We have shown
that under certain conditions there is a direct relation between these
two quantities, which is valid for resonance and off-resonance cases. Such
relation suggests that the nonclassical effects stored in the radiation
field can be noticed through the behavior of $\langle\hat{\sigma}_{z}(T)\rangle$.
Furthermore, based on this relation we have discussed the
possibility of detecting $\langle\hat{\sigma}_{z}(T)\rangle$ for single-mode
case using
techniques similar to those applied to the $W$ function.

The results given in this paper
are valid to any JCM hamiltonian  provided that it has been deduced in the framework
of rotating wave approximation.
More illustratively, if in (\ref{6}) the operators
$\prod\limits_{j=1}^{N}\hat{a}^{k_{j}}_{j}$ are replaced
by $\hat{a}^{k_{1}}_{1} \hat{a}^{\dagger k_{2}}_{2}
\hat{a}^{k_{3}}_{3} \hat{a}^{\dagger k_{4}}_{4}\cdots$, i.e. some of the annihilation
operators are replaced by creation ones,
the relation (\ref{18}) will not be affected  owing to the fact
\begin{equation}
\langle n-k|\exp(i\pi \hat{a}^{\dagger}\hat{a})|n-k\rangle\equiv
\langle n+k|\exp(i\pi \hat{a}^{\dagger}\hat{a})|n+k\rangle=(-1)^{n+k}.
\label{25}
\end{equation}
Moreover, the results are independent of the type of the
initial atomic state (, i.e. if the atom is in the excited state, ground  state or  atomic
superposition state).
Basically expression (\ref{18}) depends on the Fock state basis
of the dynamical wave function, e.g. for the single-mode case it depends on
$|n\rangle, |n+k\rangle$ and $|n-k\rangle$. Consequently, such relation
still exists when Jaynes-Cummings hamiltonian includes Kerr
nonlinearity \cite{kerr}, Stark effect \cite{stark}, intensity
dependence
\cite{intensity} and atomic motion \cite{motion}.
Furthermore, among all quasiprobability distribution functions
such relation, i.e. (\ref{18}), exists  only for the $W$ function.
 For instance, the evolution of the $Q$ function at the phase space origin
 for single-mode JCM with initial coherent light and $\triangle=0$ is given by
\begin{equation}
Q(0,T)=\exp(-2|\alpha|^{2})\cos^{2}(T\sqrt{k_{1}!}).\label{26}
\end{equation}
It is obvious that in the strong-intensity regime $Q(0,T)\rightarrow 0$.

Finally,
as it is well known, the  $W$ function  is a global quantity which
 characterizes the full quantum state.
Additionally, dealing with  the $W$ function at
 an isolated single point causes a  difficulty in
 finding a  proper normalization. Nevertheless, throughout the paper
 we have focused the attention on the behavior of WOP and its
 consistence with
 the behavior of
 $\langle \hat{\sigma}_{z}(T)\rangle$. Therefore
  the normalization   has no effect on the dynamical
  behavior of the system.

\section*{Appendix}

In this appendix we derive the asymptotic form (\ref{24b}) for $P(0,T)$.
It is worth remembering that
in the strong-intensity regime  (, i.e. $|\alpha|>>1$) the argument
 of $\cos(.)$ in (\ref{24a})
  can be expressed as \cite{rice}:

\begin{math}
\sqrt{n+1}=\sqrt{\langle \hat{n}\rangle+n+1 - \langle \hat{n}\rangle} \simeq
\frac{1}{2}(\bar{n}+
\frac{1}{\bar{n}}
+\frac{n}{\bar{n}}), \hfill(1a)
\end{math}

\noindent where $\bar{n}=\sqrt{\langle\hat{n}(0)\rangle}$.

Now we show how the different summations in (\ref{24a}) can be evaluated.

\begin{math}
\sum\limits_{n=0}^{\infty}\frac{|\alpha|^{2n}}{(n!)^{2}2^{n}}{\rm
H}^{2}_{n}(0)
=\frac{1}{2\pi}\int\limits_{0}^{2\pi}
\sum\limits_{n=0}^{\infty}\frac{(\frac{\alpha}{\sqrt{2}})^{n}}{n!}{\rm
H}_{n}(0)
\sum\limits_{m=0}^{\infty}\frac{(\frac{\alpha^{*}}{\sqrt{2}})^{m}}{m!}{\rm
H}_{m}(0)d\phi, \hfill(2a)
\end{math}

\noindent where $\alpha=|\alpha|\exp(i\phi)$.
By means of the generating function
of the Hermite polynomial the summations on the right hand side of $(2a)$ can be carried out as

\begin{math}
\sum\limits_{n=0}^{\infty}\frac{|\alpha|^{2n}}{(n!)^{2}2^{n}}{\rm
H}^{2}_{n}(0)
=\frac{1}{2\pi}\int\limits_{0}^{2\pi}
\exp[-|\alpha|^{2}\cos(2\phi)]
d\phi={\rm I}_{0}(|\alpha|^{2})
, \hfill(2b)
\end{math}

\noindent where $I_{0}(.)$ is the modified Bessel function of the first
kind of order zero.
The second summation we would like to evaluate is the following

\begin{math}
\sum\limits_{n=0}^{\infty}\frac{|\alpha|^{2n}}{n!(n+1)!2^{n+1}}{\rm
H}^{2}_{n+1}(0)
=
\sum\limits_{m=0}^{\infty}\frac{|\alpha|^{2(m-1)}}{(m-1)!m!2^{m}}{\rm
H}^{2}_{m}(0), \hfill(3a)
\end{math}

\noindent where the factorial $-1!$ is $\infty$.
Summation (3a) can be reformulated  as

\begin{math}
\sum\limits_{n=0}^{\infty}\frac{|\alpha|^{2n}}{n!(n+1)!2^{n+1}}{\rm
H}^{2}_{n+1}(0)
=\frac{d}{d|\alpha|^{2}}
\sum\limits_{m=0}^{\infty}\frac{|\alpha|^{2m}}{(m!)^{2}2^{m}}{\rm
H}^{2}_{m}(0). \hfill(3b)
\end{math}

\noindent Using  (2b), the right hand side of (3b) gives

\begin{math}
\frac{d}{d|\alpha|^{2}}{\rm I}_{0}(|\alpha|^{2})=
{\rm I}_{1}(|\alpha|^{2}). \hfill(3c)
\end{math}

The terms including $\cos(.)$  in the strong-intensity regime
 can be written as

\begin{math}
\sum\limits_{n=0}^{\infty}\frac{|\alpha|^{2n}}{(n!)^{2}2^{n}}{\rm
H}^{2}_{n}(0)
\cos
\left(T(\bar{n}+
\frac{1}{\bar{n}}
+\frac{n}{\bar{n}})\right)=
{\rm Re}\Bigl\{\exp[
iT(\bar{n}+
\frac{1}{\bar{n}})]
\sum\limits_{n=0}^{\infty}\frac{|\alpha|^{2n}}{(n!)^{2}2^{n}}{\rm
H}^{2}_{n}(0)\exp(iT\frac{n}{\bar{n}})\Bigr\}. \hfill(4a)
\end{math}

\noindent The summation on the right hand side of (4a) can be
evaluated using procedures as those given above leading that

\begin{math}
\sum\limits_{n=0}^{\infty}\frac{|\alpha|^{2n}}{(n!)^{2}2^{n}}{\rm
H}^{2}_{n}(0)
\cos
\left(T(\bar{n}+
\frac{1}{\bar{n}}
+\frac{n}{\bar{n}})\right)=
{\rm Re}\Bigl\{\exp[
iT(\bar{n}+
\frac{1}{\bar{n}})]
   {\rm I}_{0}(|\alpha|^{2}\exp(i\frac{T}{\bar{n}}))\Bigr\}. \hfill(4b)
\end{math}

\noindent Similarly  the last summation in (\ref{24a}) can be performed.

\section*{ Acknowledgement}
I would like to thank Professors  Z. Hradil (Department of Optics,
Palack\'y University, Olomouc, Czech Republic)  and  M. G.
D'Araino (Dipartimento di Fisica ``A. Volta'' via Bassi 6,
I-27100 Pavia, Italy) for the interesting discussions about
homodyne tomography. Also I thank  Professor M. R. B. Wahiddin
(Centre for Computational and Theoretical Sciences, Kulliyyah of
Science, International Islamic University Malaysia, 53100 Kuala
Lumpur, Malaysia) for his kind hospitality during my stay.

\end{document}